\documentclass[journal,12pt,onecolumn]{IEEEtran}
\pdfoutput=1
\usepackage{epsfig}
\usepackage{graphicx}

 \usepackage{setspace}
\newcommand{\beq}{\begin{equation*}}
\newcommand{\eeq}{\end{equation*}}

\usepackage{amsmath,amstext,amsfonts,latexsym,amssymb}
\usepackage{bm,times,pslatex}

\title{Spectral Efficiency of Spectrum Pooling Systems}

\author{Majed Haddad %\thanks{This paper is a preprint of a paper accepted by IET special issue on Cognitive Spectrum Access and is subject to IET copyright [URL]. When the final version is published, the copy of record will be available at [IET Digital Library URL]}
\thanks{The work reported herein was partially
supported by the projects GRACE and E2R2. This work was also
supported by Alcatel-Lucent within the Alcatel-Lucent Chair on
flexible radio at SUPELEC.
Parts of this paper were presented at GlobeCom 2007 \cite{ieeeIT.Majed2}.} and Aawatif Hayar\\
Mobile Communications Group, Institut Eurecom, \\
2229 Route des Cretes, B.P. 193,\\ 06904 Sophia Antipolis, France\\
Email: majed.haddad@eurecom.fr, aawatif.hayar@eurecom.fr
\and \\
M\'{e}rouane Debbah\\
SUPELEC, Plateau de Moulon, 3 rue Joliot-Curie\\
91192 Gif sur Yvette Cedex, France\\
Email: merouane.debbah@supelec.fr}

\begin{document}
\bibliographystyle{IEEEbib}
%\ninept

%

\maketitle

\begin{abstract}
In this contribution, we investigate the idea of using \emph{cognitive radio} to reuse
locally unused spectrum to increase the total system capacity. We consider a multiband/wideband system in which the primary and cognitive users wish to communicate to different receivers, subject to mutual interference and assume that each user knows only his channel and the unused spectrum through adequate sensing. The basic idea under the proposed scheme is based on the notion of \emph{spectrum pooling}.
The idea is quite simple: a cognitive radio will listen to the channel and, if sensed idle, will transmit during the voids. It turns out that, although its simplicity, the proposed scheme showed very interesting features with respect to the spectral efficiency and the maximum number of possible pairwise cognitive communications. %\textbf{Within this setting, we use two simple methods for sensing the idle sub-bands over the total bandwidth.}
We impose the constraint that users successively transmit over available bands through selfish water filling. For the first time, our study has quantified the asymptotic (with respect to the band) achievable gain of using spectrum pooling in terms of spectral efficiency compared to classical radio systems. We then derive the total spectral efficiency as well as the maximum number of possible pairwise communications of such a spectrum pooling system.

\end{abstract}

\vspace{1cm}
%\vspace{-.3cm}

\section{Introduction}
\label{sec:intro} The recent boom in personal wireless technologies has led to an increasing demand in terms of spectrum resources. To combat this overcrowding, the FCC has recently recommended \cite{book.FCC}
%,\cite{book.FCC2})
that significantly greater spectral efficiency could be realized by
de-ploying wireless devices that can coexist with the licensed
(primary) users, generating minimal interference while taking
advantage of the available resources. The current approach for
spectrum sharing is regulated so that wireless systems are assigned
fixed spectrum allocations, operating frequencies and bandwidths,
with constraints on power emission that limits their range.
Therefore, most communication systems are designed in order to
achieve the best possible spectrum efficiency within the assigned
bandwidth using sophisticated modulation, coding, multiple antennas
and other techniques.

On
the other hand, the discrepancy between spectrum allocation and
spectrum use suggests that this spectrum shortage could be overcome
by allowing more flexible usage of a spectrum. Flexibility would
mean that radios could find and adapt to any immediate local
spectrum availability. A new class of radios that is able to
reliably sense the spectral environment over a wide bandwidth,
detect the presence/absence of legacy users (primary users) and use
the spectrum only if the communication does not interfere with
primary users is defined by the term \emph{cognitive radio}
\cite{book.Mitola}. Cognitive radios (CR) have been proposed as a mean to
implement efficient reuse of the licensed spectrum. The key feature
of cognitive radios is their ability to recognize their
communication environment and \emph{independently} adapt the
parameters of their communication scheme to maximize the quality of
service (QoS) for the secondary (unlicensed) users while minimizing
the interference to the primary users.\\
The basic idea within the paper is based on \emph{spectrum pooling}.

The notion of spectrum pooling was first mentioned in \cite{ieeeIT.Mitola_pooling}. It
basically represents the idea of merging spectral ranges from
different spectrum owners (military, trunked radio, etc.) into a
common pool. It also reflects the need for a completely new way
of spectrum allocation as proposed in \cite{ieeeIT.Jondral_pooling}. The goal of spectrum pooling is to enhance spectral efficiency
by overlaying a new mobile radio system on an existing one
without requiring any changes to the actual licensed system.

Another technique that has been increasingly popular is Time
Division Duplexing (TDD) in which the same carrier is used for both
links in different time slots. One property of such systems is that,
since the same frequency is used, the channel characteristics are
nearly the same in both links, provided the channel does not change
too rapidly.

Motivated by the desire for an effective and practical scheme, our study treats the
problem of spectrum pooling from sensing to achievable performance. We consider an asynchronous TDD communication scenario in which the
primary and cognitive users wish to communicate to different
receivers, subject to mutual interference in a heterogeneous network
where devices operates in a wideband/multiband context. However,
contrary to the work addressed in \cite{ieeeIT.Majed}, in this
contribution, we impose as a first step that only one user can simultaneously
transmit over the same sub-band using successive water filling. Especially OFDM based WLANs like IEEE802.11a and
HIPERLAN/2 are suitable for an overlay system like spectrum pooling as they allow a very flexible frequency management on a carrier-by-carrier basis. We examine the total spectral
efficiency of the spectrum pooling system and show that the overall
system spectral efficiency can be considerably enhanced by
considering cognitive communications with respect to the traditional
system (without cognition). In particular, it is of major interest,
in this context, to quantify the spectral efficiency gain in order
to show the interest behind using spectrum pooling terminals with
respect to classical systems (without
cognition). In fact, although spectrum polling
have spurred great interest and excitement, many of the
fundamental theoretical questions on the limits of such technologies
remain unanswered.
The merits of our approach lie in the simplicity of the proposed scheme and, at the same time, its efficiency. Results showed very interesting performance in terms of the number of cognitive users allowed to transmit as well as the system spectral efficiency gain we get.
Such an accurate and simple system modeling presents a key to understand the actual benefits brought by spectrum pooling technology.\\\\
The rest of the paper is organized as follows: In Section II, we
describe the channel model. In Section III, we describe the spectrum pooling protocol. In Section IV, we address the problem of sensing. Section V details the
spectral efficiency analysis adopted throughout this paper when the
number of sub-bands is limited. In Section VI, we investigate the
asymptotic performance of such a system in terms of spectral efficiency. Performance evaluation is provided in Section VII
and Section VIII concludes the paper.
\section{The channel model}
The baseband discrete-frequency model at the receiver $\mathfrak{R}_{l}$ %with $L$ users and $N$ sub-bands
(see Figure \ref{fig:system}) is:
\begin{equation}
y^i_{\mathfrak{R}_{l}} = \displaystyle h_l^i \sqrt{P_l^i(h_{l}^i)} S_l^i + n_l^i,  \qquad \mathrm{for}\,\,\,\ i = 1,...,N \qquad \mathrm{and} \qquad l = 1,...,L
\label{eq:sysfreq}\end{equation}
where:\begin{itemize}
        \item  $h_l^i$: is the block fading process of user $l$ on the sub-band $i$,\vspace{0.1cm}
        \item  $S_l^i$: is the symbol transmitted by user $l$ on the sub-band $i$,\vspace{0.1cm}
        \item  $P_l^i(h_{l}^i)$: is the power control\footnote{Throughout the rest of the paper, we will find it
convenient to denote by $P_{l}^i$ the power allocation policy of user $l$ on sub-band $i$, rather than $P_{l}^i(h_{l}^i)$.\\} of user $l$ on the sub-band $i$,\vspace{0.1cm}
        \item  $n_l^i$: is the additive Gaussian noise at the $i$th sub-band.
      \end{itemize}
We further assume that the channel $h_{l}$ stays constant over each block fading
length (i.e. coherent communication). The assumption of coherent
reception is reasonable if the fading is slow in the sense
that the receiver is able to track the channel variations. We statistically model the
channel gains $h_{l}$ to be i.i.d distributed over the $L$ Rayleigh fading
coefficients and $\mathbb{E}\left\{\left|h_{l}\right|^{2}\right\}=1$ for $l = 1,...,L$. The additive Gaussian noise $n_l$ at the receiver is i.i.d circularly symmetric and $n_l\sim\mathcal{CN}(0,N_0)$) for $l = 1,...,L$.

\section{The Spectrum pooling Protocol}
\begin{figure}[http]
\begin{center}
\vspace{-2cm}\centerline{\includegraphics[width=20cm,height=16cm]{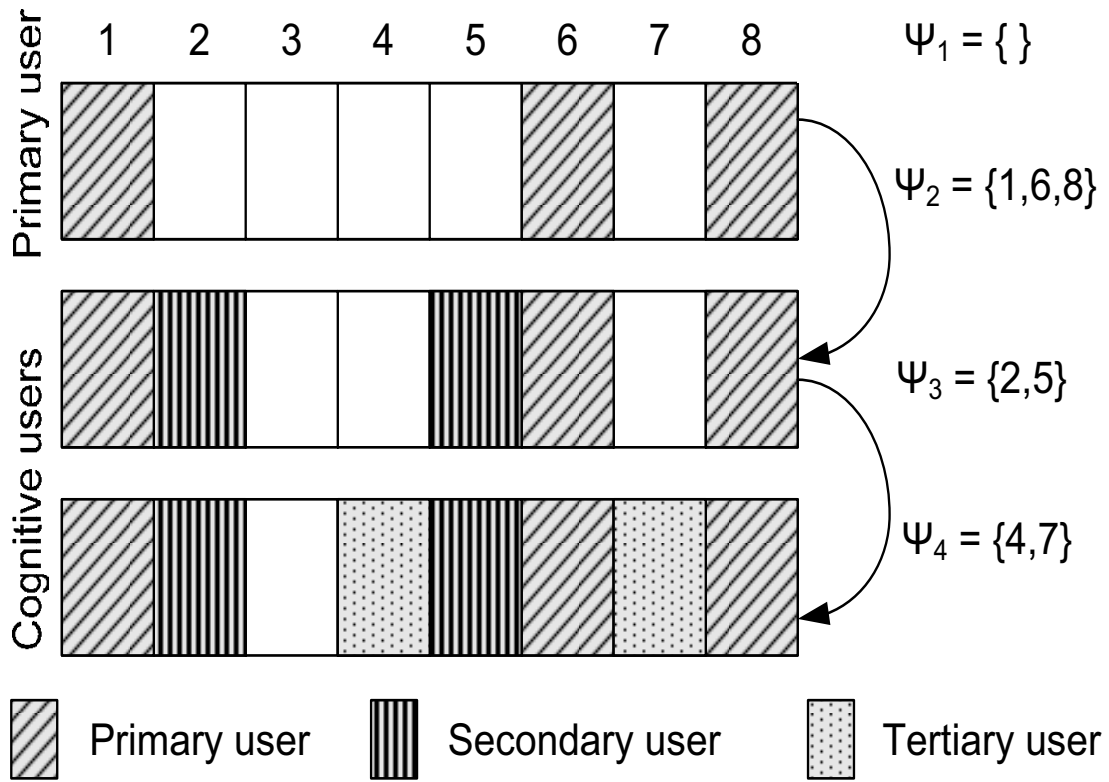}}
\vspace{-0.5cm} \caption{One primary user and two cognitive users in
a system with 8 sub-bands. } \label{fig:band}\end{center}
\end{figure}

We consider an asynchronous TDD communication scenario in which the primary and cognitive users
wish to communicate to different receivers, subject to mutual interference. The basic idea under the proposed protocol is quite simple: the cognitive users listen to the wireless
channel and determine, either in time or frequency, which part of
the spectrum is unused. Then, they \emph{successively} adapt their
signal to fill detected voids in the spectrum domain. Each
transmitter $\mathcal{T}_l$ for $l = 1,...,L$ estimates the pilot
sequence of the receiver $\mathfrak{R}_{l}$ in order to determine
the channel gain $h_l$ (see links (1) and (3) in Fig.
\ref{fig:system}). Notice here that since we are in a TDD mode, when
we estimate the channel in one way, we can also know it the other
way. Thus, each user $l$ is assumed to know only his own channel
gain $h_l$ and
the statistical properties of the other links (probability distribution). We further assume that the channel does not change from the instant of estimation to the instant of transmission.\\
A particularly noteworthy target in this context, when we employ a "\emph{listen-before-talk}" strategy, is to
reliably detect the sub-bands that are currently accessed by a specified user in order to be spared from the coming users transmission. This knowledge can be obtained from two manners: In a centralized mode where the proposed
system would require information from a third party (i.e. central database maintained by regulator
or another authorized entity) to schedule users coming. Alternatively, an extra signalling channel is dedicated to perform the collision detection so that cognitive users
%come after in an asynchronous way so that they
will not transmit at the same
moment. Specifically, the primary user comes first in the system and
estimates his channel gain. The second user comes in the system randomly,
for instance in a Poisson process manner, and estimates his
channel link. Such an assumption could be further justified by the fact that in an asynchronous context, the
probability that two users decide to transmit at the same moment is
negligible as the number of users is limited. Thus, within this setting, the primary user is assumed not to be aware of the
cognitive users. He communicates with his receiver in an
ad-hoc manner while a set of spectrum pooling transmitters that are
able to reliably sense the spectral environment over a wide
bandwidth,
%detect the presence/absence of legacy users and
decide to communicate with theirs respective receivers only if the communication does not
interfere with the primary user. Accordingly, under our opportunistic
approach, a device transmits over a certain
sub-band only when no other user does. Such an assumption is motivated by the fact that when $\mathfrak{R}_l$ sends his pilot sequence to $\mathcal{T}_l$, he will not interfere with $\mathcal{T}_{l-1}$ for $l = 2,...,L$. The sensing operation will be discussed in the next section. Throughout the rest of the paper, we will adopt this framework to analyze the achievable performance of such a system in terms of spectral efficiency as well as the maximum number of possible pairwise communication within this scenario. Such an accurate and simple system modeling presents a key to understand the actual benefits brought by spectrum pooling technology. In fact, although cognitive radios have spurred great interest and excitement in industry, many of the fundamental theoretical questions on the limits of such technologies remain unanswered.\\
Moreover, in order to characterize the achievable performance limit of such systems, three capacity measures can be found in the literature. A comprehensive review of
these concepts can be found in \cite{ieeeIT.Biglieri}.
The relevant performance metric of the proposed protocol is the instantaneous
capacity per sub-band in bits/s/Hz, also called \emph{spectral efficiency}, namely \cite{book.tse05}:
\begin{equation}
C_{l} = \frac{1}{N}\sum_{i=1}^{N}\log_2\left(1+\frac{ P_{l}^i \mid h_{l}^i
\mid^2}{N_0}\right); \qquad l = 1,...,L
\label{eq:hcarr151}\end{equation}
The sum here is done over the stationary instantaneous distribution
of the fading channel on each user $l$. The instantaneous capacity
determines the maximum achievable rate over all fading states
without a delay constraint. In this work, we allocate transmit
powers for each user (over a total power budget constraint) in order
to maximize his transmission rate. In fact, when channel state
information is made available at the transmitters, users know their
own channel gains and thus they will adapt their transmission
strategy relative to this knowledge. The corresponding optimum power
allocation is the well-known \emph{water filling} allocation
\cite{book.Cover} expressed
by\footnote{$(x)^+=\textrm{max}(0,x)$.}:
\begin{equation}
P_{l}^i =\left(
\frac{1}{\gamma_{0}}-\frac{N_0}{\left|h_{l}^i\right|^{2}}\right)^+\\
\label{eq:snrprob43}\end{equation} where $\gamma_{0}$ is the
Lagrange's multiplier satisfying the average power
constraint per sub-band:
\begin{equation}
\frac{1}{N}\sum_{i=1}^{N} P_{l}^i = \overline{P}
\label{eq:hcarr159}\end{equation}
Without loss of generality, throughout the rest of the paper, we
take $\overline{P} = 1$.\\
Notice that, although a water filling power allocation strategy is adopted in this analysis, we emphasize that this is not a restriction of the proposed protocol. In fact, as mentioned before, one
important task when implementing spectrum pooling is that cognitive users operate on the idle sub-bands of the licensed system delivering
a binary channel assignment as shown in Fig. \ref{fig:band}. Hence, our study is valid for any binary power control without resorting to the restriction assumption of successive water filling.

For clarity sake, let us take the following example with $N=8$
sub-bands.
As shown in Figure \ref{fig:band}, the primary user is always
prioritized above cognitive users by enjoying the entire band
while cognitive users adapt their signal to fill detected voids
with respect to their order of priority. As a first step, the primary user
maximizes his rate according to his channel process. As mentioned
before in expression (\ref{eq:snrprob43}), only user with a channel
gain $h^i$ above a certain threshold equal to $\gamma_{0}.N_0$
transmits on the sub-band $i$ ($\Psi_2$). User 2, comes in the system randomly, senses the spectrum and decides
to transmit only on sub-bands sensed idle. Thus, following his
fading gains, user 2 adapts his signal to fill these voids in the
spectrum domain in a complementary fashion ($\Psi_3$). Similarly, user 3 will sense the remaining
sub-bands from user 1 and user 2 and
decides to transmit during the remaining voids ($\Psi_4$). \\

\section{Sensing issue}
\begin{figure}[h!]
\begin{center}
\vspace{5cm}
\rotatebox{0}{
\includegraphics[height=14cm]{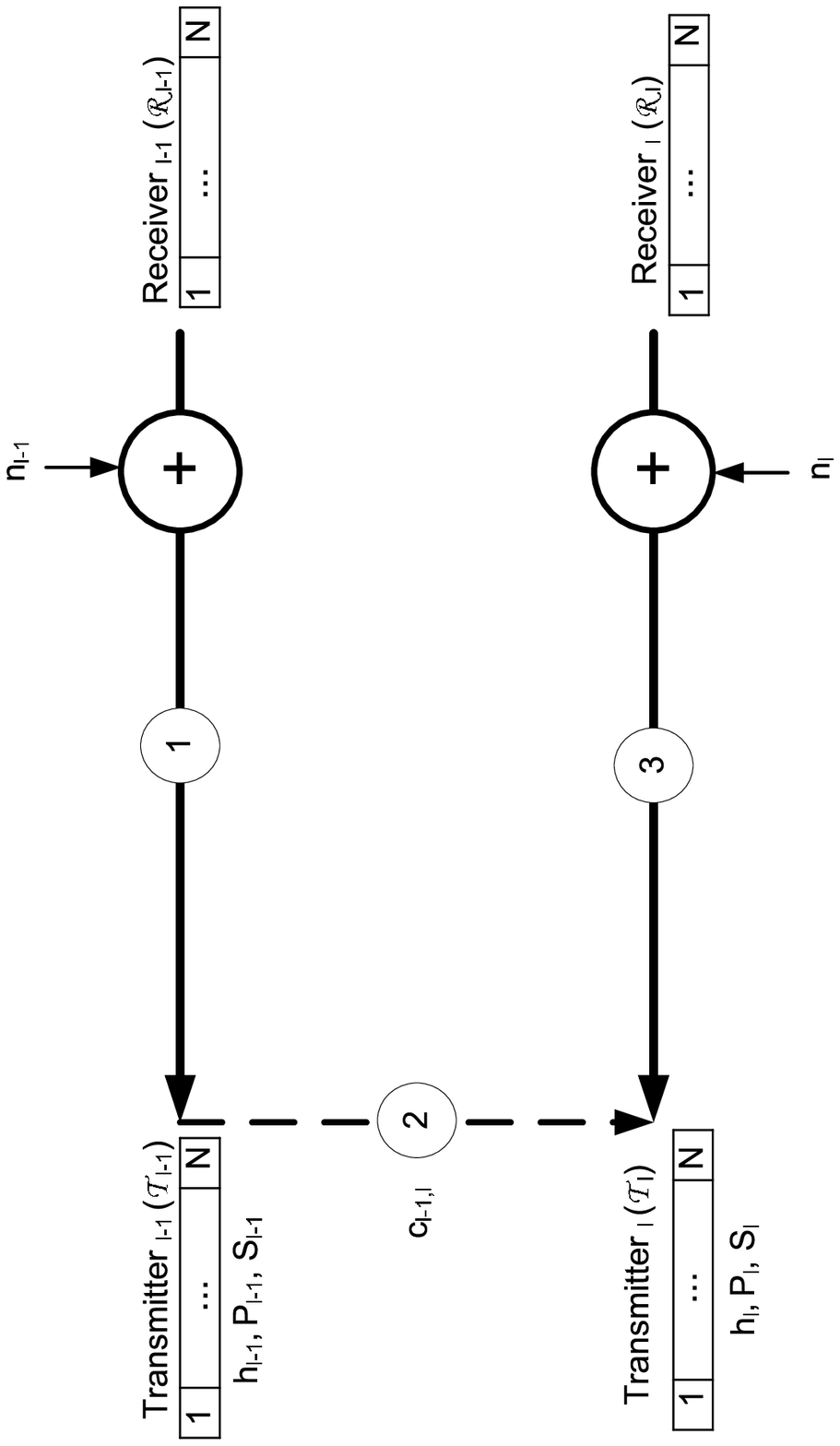}}
\vspace{0cm}
\caption{The cognitive radio channel in a
wideband/multiband context with $N$ sub-bands.}
\label{fig:system}\end{center}
\end{figure}

So far, we have focused on pairwise communications between transmitters and receivers (see links 1 and 3 in Fig. \ref{fig:system}). Let us now investigate the \emph{inter-transmitter} communications (link 2 in Fig. \ref{fig:system}) in order to analyze the problem of sensing. To this effect, let us assume the baseband discrete-time model within a coherence time period $T$ when each user $l$ for $l = 2,...,L$ has $N$ sub-bands as described in Figure \ref{fig:system}:\vspace{.4cm}
\begin{equation}
y_{l}^i(k) = c_{l-1,l}^{i}(k) \sqrt{P_{l-1}^i(h_{l-1}^i)} S_{l-1}^i(k) + n_{l-1}^i(k),
\label{eq:sysfreq}\end{equation}
where $c_{l-1,l}^{i}(k)$ is the block fading process from user $l-1$ to user $l$ on the $i$th sub-band, at time $k$. We further assume that $0 \leq k \ll \beta T$ and $\beta < 1$, i.e. the coherence time is sufficiently large so
that the channel stays constant for samples and
jumps to a new independent value (block-fading model).\\
%$\mathcal{T}_{l-1}$ sends a pilot sequence to $\mathcal{T}_l$.
The proposed sensing techniques hinge on the assumption that all devices operate under a unique standard so that they know the pilot sequence used by the other users.

\begin{figure}[http]
\begin{center}
\includegraphics[width=14cm,height=11cm]{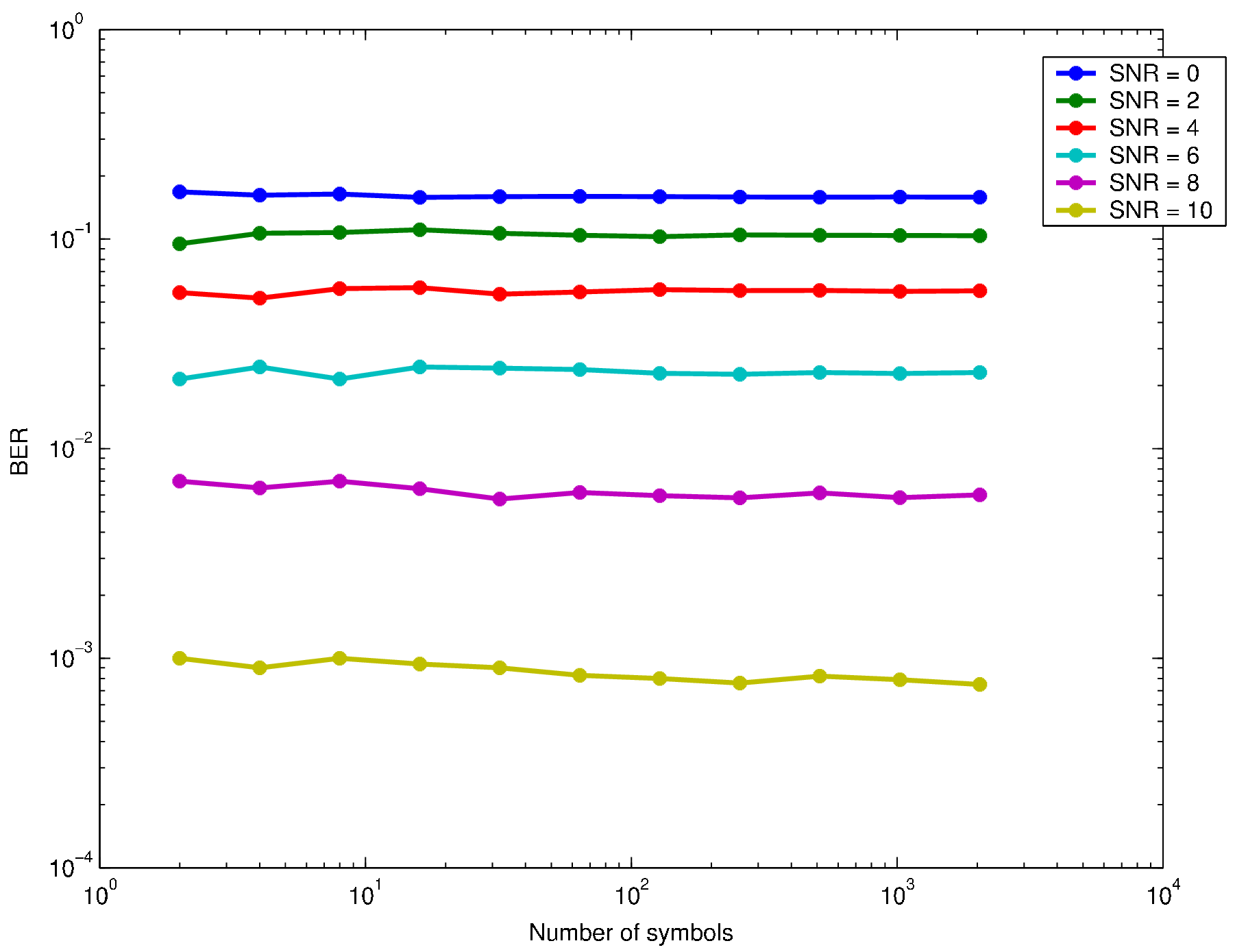}
\vspace{1cm}
\caption{BER v.s number of symbols ($M$) in dB for BPSK in AWGN
using power detection where SNR are in dB.}
\label{fig:cap6}\end{center}
\end{figure}

As stated above, in this work, the spectrum pooling behavior is
assumed to allow only one user to simultaneously transmit over
the same sub-band. The received signal at user $l$ can therefore be
written as (see link 2 in Fig. \ref{fig:system}):
%\textbf{think about changing S for symbol}
\vspace{.4cm}
\begin{equation}
y_{l}^i(k)  =\left\{ \begin{array}{l} \underbrace{c_{l-1,l}^{i}(k)
\sqrt{P_{l-1}^i} S_{l-1}^i(k)} + \underbrace{n_{l-1}^i(k)}, \qquad
\mathrm{if}\,\,\,\ P_{l-1}^i \neq 0  \\ \qquad \qquad \mathrm{signal} \qquad \qquad \mathrm{noise}\\
n_{l-1}^i(k),\,\,\,\qquad\qquad\qquad\qquad\qquad\qquad\mathrm{otherwise}\end{array}\right.
\label{eq:snrprob4}\end{equation}
\vspace{.6cm}
By assuming that $\beta T$ is an integer equal to $M$ and by
making $\beta T$ sufficiently large, the mean received power over the detection duration at
receiver $\mathfrak{R}_l$ is:
\begin{equation}
\lim_{M \rightarrow \infty}\frac{1}{M} \sum_{k=1}^{M} \left|y_{l}^i(k)\right|^2 = \left\{ \begin{array}{l} \left|c_{l-1,l}^i\right|^{2} P_{l-1}^i + N_0 ,\,\,\,\qquad\qquad
\mathrm{if}\,\,\,\
P_{l-1}^i \neq 0  \\\\
N_0 ,\,\,\,\qquad\qquad\qquad\qquad\qquad\mathrm{otherwise}\end{array}\right.
\end{equation}
Accordingly, in order to determine which
part of the spectrum is unused, cognitive user has just to detect
the received power and compare it to the noise power $N_0$.
However, in addition to the fact that it supposes that $M \rightarrow \infty$ (i.e. infinite time coherence period), the proposed method would be not efficient at low SNR-regime (see Figure \ref{fig:cap6}). In fact, the quality of such a technique is strongly degraded with the reduction in the precision of the noise threshold \cite{ieeeIT.Cai}\cite{book.Van}. The principal difficulty of this detection is to obtain a good estimation of the noise variance.
In the setting of spectrum pooling mechanism, we would need a channel sensing method that continuously senses the channel.
Thus, the channel sensing should be performed with a very high
probability of correct detection (to assure very low probability
of interference with the primary system). Weiss \emph{et al.} proposed in \cite{ieeeIT.Jondral_signaling_pooling} a distributed spectrum pooling protocol where all the nodes participate in channel sensing so that all cognitive users perform detection. Moreover, formulas for the calculation of the detection
and false alarm probability in a spectrum pooling system
have been derived in \cite{ieeeIT.Jondral_detection_pooling} for the general case of an arbitrary primary system's
covariance matrix.

\section{Spectral efficiency analysis}

Let us first define the set of the number of sub-bands sensed
occupied by user $l$ by: \begin{equation} \Psi_l = \left\{i \in \{ 1,...,N\}; \,\,\,P_{l-1}^i \neq 0\right\}\label{eq:snrprob124}\end{equation} where
$\Psi_l$ obeys to the following properties:
\begin{equation} \left\{\begin{array}{l}
\Psi_1 = {\o},\\\\
\displaystyle\bigcup_{l=1}^{L+1} \Psi_l \subseteq \{1,...,N\},\\\\
\displaystyle\bigcap_{l=1}^{L+1} \Psi_l = {\o}\\\\
\end{array}\right.
\label{eq:snrprob124}\end{equation} The spectral efficiency per sub-band of user
$l$, given a number of sub-bands $N$, is:
\begin{equation}
C_{l,N} = \frac{1}{card(\Omega_l)}\displaystyle\sum_{i\in \Omega_l}
\log_2\left(1+ \frac{P_{l}^i \mid h_{l}^i
\mid^2}{N_0}\right)\qquad bits/s/Hz
\label{eq:sytfeqz}\end{equation} where $\Omega_l$ represents the set
of the remaining idle sub-bands sensed by user $l$, namely:
\begin{equation}
\Omega_l = \left\{i \in \{1,...,N\} \bigcap
\displaystyle\overline{\bigcup_{k=1...l}\Psi_k}   \right\}
\label{eq:sfeq}\end{equation} For a given number of sub-bands $N$,
the optimal power allocation which maximizes the transmission rate
of user $l$ is the solution to the following optimization problem:
\begin{equation*}
\max_{P_{l}^1,...,P_{l}^{card(\Omega_l)}} C_{l,N},\qquad
\mathrm{for}\,\,\,\  l=1,...,L\end{equation*}
\\subject to the average power constraint per sub-band: \begin{equation}
\left\{\begin{array}{l} \displaystyle \frac{1}{card(\Omega_l)}
\displaystyle \sum_{i\in \Omega_l}  P_{l}^i = 1,\\\\P_{l}^i
\geq 0,\\\\
\end{array}\right.
\label{eq:snrprob124}\end{equation} The resulting optimal power
control policy is given by (\ref{eq:snrprob43}).
%The short-term sum
%capacity over the system is given by :
%\begin{equation}
%C_{sum,N} = \sum_{l=1}^{L} C_{l,N} \label{eq:sfeq}\end{equation}
Notice that the maximum number of users $L$ allowed by such a system
must satisfy the
condition that $card(\Omega_L)\neq0$.\\
Let us now derive the spectral efficiency of such a system. The
spectral efficiency per band of user $l$ is given by:
\begin{equation}
\Phi_{l,N} \begin{array}[t]{l} = \displaystyle \frac{1}{N} \cdot
\displaystyle\sum_{i\in \Omega_l} \log_2\left(1+ \frac{P_{l}^i \mid
h_{l}^i \mid^2}{N_0}\right)
\end{array}\,\,\label{eq:sytfeq}\end{equation}
By multiplying and dividing (\ref{eq:sytfeq}) by $card(\Omega_l)$,
we obtain\footnote{Notice that since the primary user enjoys
the entire bandwidth, we have: $card(\Omega_1)=N$.}:
\begin{equation} \Phi_{l,N}= %\left\{\begin{array}{l}
%C_{1,N}, \qquad\qquad \qquad \mathrm{if}\,\,\,\ l = 1\\\\
\displaystyle \frac{card(\Omega_l)}{N} . C_{l,N},\qquad
\mathrm{for}\,\,\,\  l=1,...,L
%\end{array}\right.\,\,
\label{eq:syfe7q}\end{equation}
As expected, when $l=1$, the spectral efficiency without cognition is equal to the primary user spectral efficiency $C_{1,N}$.
We define $\Delta_{l,N}$ as the band factor gain of user $l$ for $N$
sub-bands, namely:
\begin{equation}
\Delta_{l,N} \triangleq \frac{card(\Omega_l)}{N},\qquad
\mathrm{for}\,\,\,\
l=1,...,L\label{eq:sywfequ}\end{equation} In other words,
the band factor gain represents the fraction of the band unoccupied
at user $l$. The spectral efficiency per band of user $l$ can
therefore be expressed by:
\begin{equation} \Phi_{l,N}= %\left\{\begin{array}{l}
%\Delta_{1,N}. C_{1,N}, \qquad\qquad \mathrm{if}\,\,\,\ l = 1\\\\
\displaystyle \Delta_{l,N} \cdot C_{l,N},\qquad \mathrm{for}\,\,\,\
l=1,...,L
%\end{array}\right.\,\,
\label{eq:syfeq}\end{equation}
and the sum spectral efficiency of a system with $N$ sub-bands per
user is given by:
\begin{equation}
\Phi_{sum,N} = \sum_{l=1}^{L} \Phi_{l,N}
\label{eq:syfteq}\end{equation}

\section{Asymptotic Performance}

Let us now study the achievable performance when devices operate in
a wide-band context (i.e. $N\rightarrow\infty$). The spectral efficiency of user $l$ for a large number of sub-bands in
(\ref{eq:sytfeqz}) becomes:
\begin{equation}
C_{l,\infty} = \displaystyle \int_{0}^{\infty}\log_2 \left( 1+\frac{
P_{l}(t)\cdot t}{N_0}\right)\cdot f(t) dt,\,\,\,\ \mathrm{for}\,\,\,\
l=1,...,L\label{eq:hcarc45}\end{equation}
where $P_{l}$
is subject to the average constraint:
\begin{equation}\displaystyle  \int_{0}^{\infty} P_{l}(t) \cdot f(t)dt =
1\label{eq:hcarh}\end{equation}
Although this is not a restriction of our approach, from now on we assume that the
channel gains are i.i.d Rayleigh distributed. However, all theoretical results as well
as the methodology adopted in this paper can be translated
immediately into results for any other probability distribution
function of the channel model. In this way, the term $f(t)$ in
(\ref{eq:hcarc45}) will be replaced by the appropriate probability
distribution function. The spectral efficiency of user $l$ for i.i.d Rayleigh fading is given by:
\begin{equation}
C_{l,\infty} = \displaystyle \int_{0}^{\infty}\log_2 \left( 1+\frac{
P_{l}(t)\cdot t}{N_0}\right)\cdot e^{-t}dt,\,\,\,\ \mathrm{for}\,\,\,\
l=1,...,L\label{eq:hcarc4785}\end{equation} where $P_{l}$
is subject to the average constraint:
\begin{equation}\displaystyle  \int_{0}^{\infty} P_{l}(t) \cdot e^{-t}dt =
1\label{eq:hcarh}\end{equation}
and $\gamma_{0}$ is the Lagrange's
multiplier satisfying\footnote{$E_{i}(x) $ is the exponential
integral function defined as:
$E_{i}(x)=\int^{+\infty}_{x}\frac{e^{-t}}{t}dt.$}:
\begin{equation}
\frac{1}{\gamma_{0}}\displaystyle
\int^{+\infty}_{\gamma_0\cdot N_0}e^{-t}dt - N_0\cdot
E_i\left({\gamma_{0}\cdot N_0}\right) = 1
\label{eq:snrprob101}\end{equation}
Numerical root finding is needed to determine different values of
$\gamma_{0}$. Our numerical results, in section VII, show that
$\gamma_{0}$ increases as $N_0$ decreases, and $\gamma_{0}$
always lies in the interval [0,1]. On the other hand, an asymptotic
expansion of (\ref{eq:snrprob101}) in \cite{book.Gradshteyn}
shows that at very high \textrm{SNR}-regime, $\gamma_{0}\rightarrow
1$.\\
Moreover, the spectral efficiency of user
$l$ can be computed for $l=1,...,L$ as follows:
\begin{equation}
C_{l,\infty} \begin{array}[t]{l} = \displaystyle
\int_{0}^{\infty}\log_2 \left( 1+\frac{
P_{l}(t)\cdot t}{N_0}\right)\cdot e^{-t}dt\\\\

= \displaystyle \int_{\gamma_{0}N_0}^{\infty}\log_2 \left( 1+
\frac{\left(
\frac{1}{\gamma_{0}}-\frac{N_0}{t}\right)\cdot t}{N_0}
\right)\cdot e^{-t}dt\\\\

= \displaystyle \int_{\gamma_{0}N_0}^{\infty}\log_2 \left(
\frac{t}{\gamma_{0}\cdot N_0}\right) \cdot e^{-t}dt\\\\

= \displaystyle
\frac{1}{\ln(2)}\cdot E_i\left({\gamma_{0}\cdot N_0}\right)
\end{array}\,\,
\label{eq:hcaror12}\end{equation}
In order to characterize the achievable performance of such system
in terms of spectral efficiency, we define the spectral efficiency within the frequency bandwidth W, as \cite{ieeeIT.Verdu}:
\begin{equation}
C_{l,\infty} (W) = \frac{1}{W}\displaystyle \int_{\frac{-W}{2}}^{\frac{W}{2}}\log_2
\left( 1+\frac{ P_l(f).\left|H_l(f)\right|^{2}}{N_0}\right)df\\\\
\label{eq:hcaror1255}\end{equation} By
identifying expression (\ref{eq:hcarc4785}) with (\ref{eq:hcaror1255}),
we obtain a characterization of the frequency variation $f$ as
function of the channel gains $t$, namely:
\begin{equation}
f = -W \cdot e^{-t} + \frac{W}{2},\qquad %\mathrm{for}\,\,\,\
%l=1,...,L
\label{eq:hcaror1002}
\end{equation}
Similar to our approach in the previous section, we define the band factor
gain $\Delta_{\infty}$ as the fraction of the band sensed idle from user $l$
to user $l+1$ over the total bandwidth $W$ for an infinite number of sub-bands:
\begin{equation}
\Delta_{\infty} \triangleq  \frac{\Delta f}{W}%\qquad \mathrm{for}\,\,\,\
\end{equation}
where $\Delta f$ represents the frequency interval
where the fading gain in (\ref{eq:hcaror1002}) is below a certain threshold equal to
$\gamma_{0}\cdot N_0$. By deriving the appropriate vacant band $\Delta f$ when $t \in [0,\gamma_{0}\cdot N_0]$ in (\ref{eq:hcaror1002}), we obtain:
\begin{equation}
\Delta_{\infty}= 1- exp\left(-\gamma_{0}\cdot N_0\right)
%,\qquad \mathrm{for}\,\,\,\  l=2,...,L
\label{eq:hcaror2131}
\end{equation}
Accordingly, the asymptotic spectral efficiency of user $l$ is
given by:
\begin{equation} \Phi_{l,\infty}= %\left\{\begin{array}{l}
%C_{1,\infty}, \qquad\qquad \mathrm{if}\,\,\,\ l = 1\\\\
\displaystyle \Delta_{\infty}\cdot C_{l,\infty},\qquad
\mathrm{for}\,\,\,\  l=1,...,L
%\end{array}\right.\,\,
\label{eq:syfeq}\end{equation}
Similar to the case where the number of sub-bands is fixed, when $l=1$, the spectral efficiency without cognition is equal to the primary user spectral efficiency $C_{1,\infty}$. In particular, it is of major interest to quantify the spectral efficiency gain $\Delta_{\infty}$ in order to show  the interest behind using spectrum pooling terminals with respect to classical systems (without cognition).
To do so, following the same procedure and going from user 2 to $L$, we obtain
the expression of the asymptotic spectral efficiency as function
of $C_{1,\infty}$:
\begin{equation}
%\Phi_{l,\infty} =  \Delta_{l,\infty}. C_{1,\infty},\qquad
\Phi_{l,\infty} =  \Delta_{\infty}^{l-1}.
C_{1,\infty},\qquad\mathrm{for}\,\,\,\  l=1,...,L
\label{eq:syfe3q}\end{equation}
The overall asymptotic sum spectral efficiency for a system with
$L$ users is therefore:
\begin{equation}
\Phi_{sum,\infty}\begin{array}[t]{l}
= \displaystyle \sum_{l=1}^{L}\Phi_{l,\infty}\\\\
= \displaystyle \sum_{k=0}^{L-1} \Delta_{\infty}^k C_{1,\infty}\\\\
%= \displaystyle \sum_{k=0}^{L-1} \Delta^k C_{1,\infty}\\\\
= \underbrace{\frac{1-\Delta_{\infty}^L}{1-\Delta_{\infty}}}\cdot
C_{1,\infty}\\\qquad\geq 1
\end{array}\,\,
\label{eq:hcaror213}\end{equation}
Thus, the sum spectral efficiency obtained by considering cognitive
communications is greater than or equal to the spectral efficiency
without cognition $C_{1,\infty}$. Such a result, rather intuitive, justifies the
increasing interest behind using cognitive radio terminals in future
wireless communication systems since the sum spectral efficiency of
such systems performs always better than classical communication
systems (without cognition).\\
On the other hand, by substituting $C_{1,\infty}$ by
its expression in (\ref{eq:hcaror12}), we obtain the final
expression of the achievable sum spectral efficiency in such a
system:
\begin{equation}
\Phi_{sum,\infty} =
\frac{1}{\ln(2)}\cdot \frac{1-\Delta_{\infty}^L}{1-\Delta_{\infty}}\cdot
%\frac{1}{\ln(2)}\cdot \frac{1-\Delta^L}{1-\Delta}.
E_i\left(\gamma_{0}\cdot N_0\right)
\label{eq:hcaror262}\end{equation} This result is very interesting
as, by only knowing the statistics of the channel gains (through
$\gamma_0$) and the $\textrm{SNR}$ (through $N_0$), one can derive the achievable
spectral efficiency as well as the potential gain resulting from using spectrum pooling.

\section{Performance evaluation}
\begin{figure}[http]
\begin{center}
\vspace{-.7cm}
\centerline{\includegraphics[width=14cm,height=11cm]{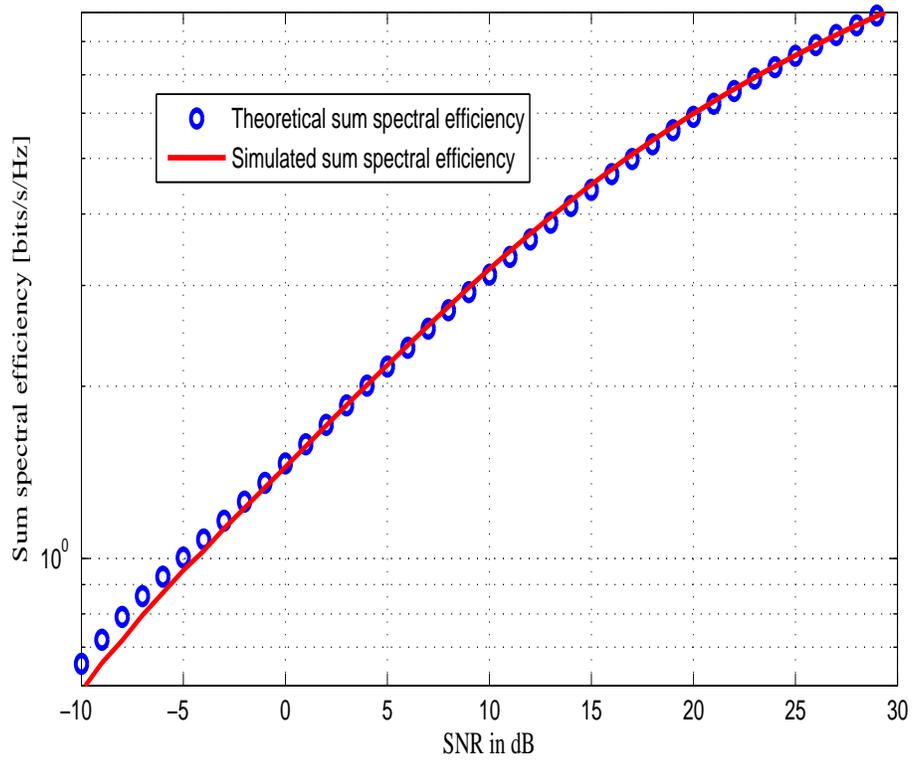}}  \vspace{.5cm}
\caption{Comparison between theoretical expression of the sum
spectral efficiency in (\ref{eq:hcaror262}) and simulated one in
(\ref{eq:syfteq}) for $L=5$ and $N$=16.}
\label{fig:cap3}\end{center}
\end{figure}

In order to validate our approach in the previous Section, we compare the theoretical expression of the sum spectral efficiency  in (\ref{eq:hcaror262})
to expression in (\ref{eq:syfteq}). We model $L$ i.i.d Rayleigh channels (one for each user) and assume
perfect sensing of the idle-sub-bands. Our numerical result in Figure \ref{fig:cap3}, tends to validate the asymptotic analysis we adopt throughout the paper. It clearly shows that the sum spectral efficiency in (\ref{eq:syfteq}) matches expression (\ref{eq:hcaror262})
even for a moderate number of sub-bands $N$ (from $N$ = 16).

\begin{figure}[http]
\begin{center}
%\vspace{-.2cm}
\centerline{\includegraphics[width=14cm,height=11cm]{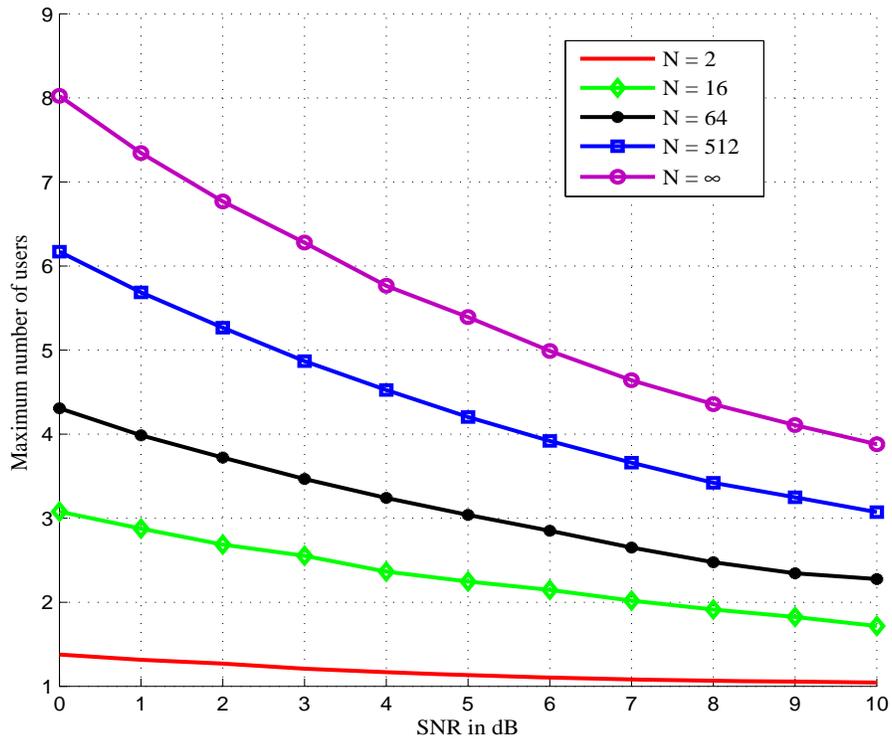}}
\vspace{1cm} \caption{The maximum number of users for different
number of sub-bands ($N$).} \label{fig:cap1}\end{center}
\end{figure}

Moreover, since the maximum number of users is not theoretically
limited, we will consider only  $L$ that satisfies the condition
that $card(\Omega_L)\neq0$, otherwise, the $L$-th spectral efficiency would be
negligible. Figure \ref{fig:cap1} characterizes the maximum number
of users $L$ as function of the received signal energy per
information bit $E_b/N_0$ for different number of sub-bands $N$.
As
expected, we remark that the maximum number of users allowed to
transmit increases with the number of sub-bands especially at low
$E_b/N_0$ region. Furthermore, the maximum number of cognitive users
ranges from 1 to 8. As an example, the proposed scheme, although its simplicity allows up to
4 cognitive users to benefit from the licensed spectrum at 8 dB for
$N = 2048$ sub-bands.\\

\begin{figure}[http]
\begin{center}
%\vspace{-1.7cm}
\centerline{\includegraphics[width=18cm,height=14cm]{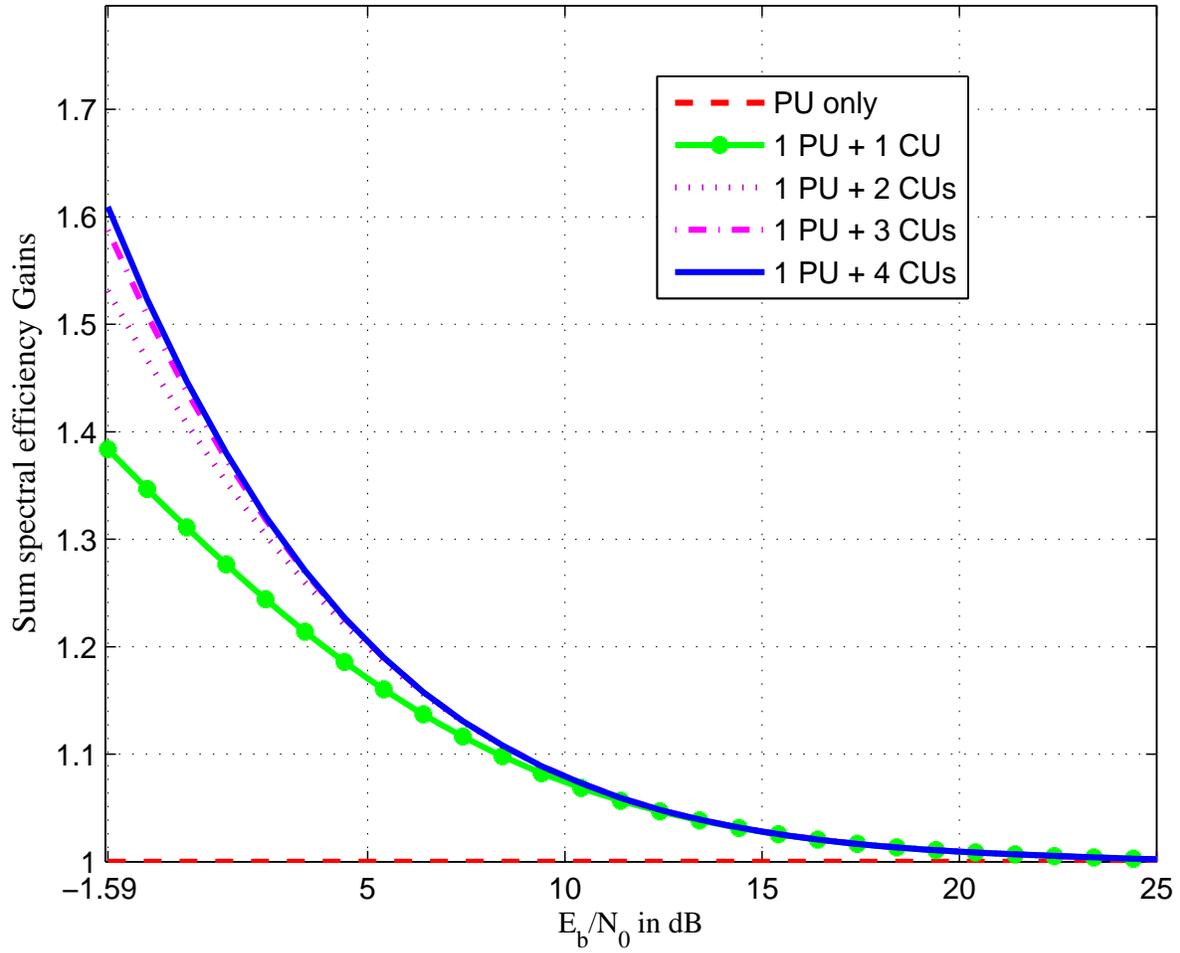}}
\vspace{-0cm} \caption{Sum spectral efficiency gains of the system with one primary user and 4 cognitive users (CU).} \label{fig:cap5}\end{center}
\end{figure}

\begin{figure}[http]
\begin{center}
%\vspace{-1.7cm}
\centerline{\includegraphics[width=14cm,height=11cm]{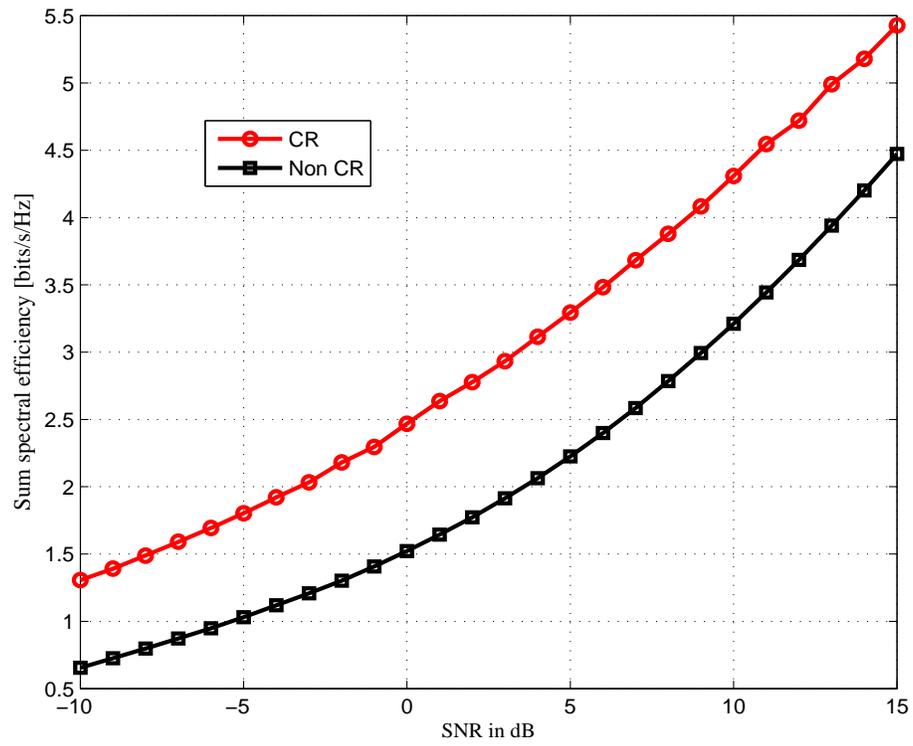}}
\vspace{0.5cm} \caption{Sum spectral efficiency of a system using cognitive radio (CR) and a traditional system (Non CR) for $N=512$.} \label{fig:cap12}\end{center}
\end{figure}

\begin{figure}[http]
\begin{center}
\vspace{0cm}
\centerline{\includegraphics[width=20cm,height=14cm]{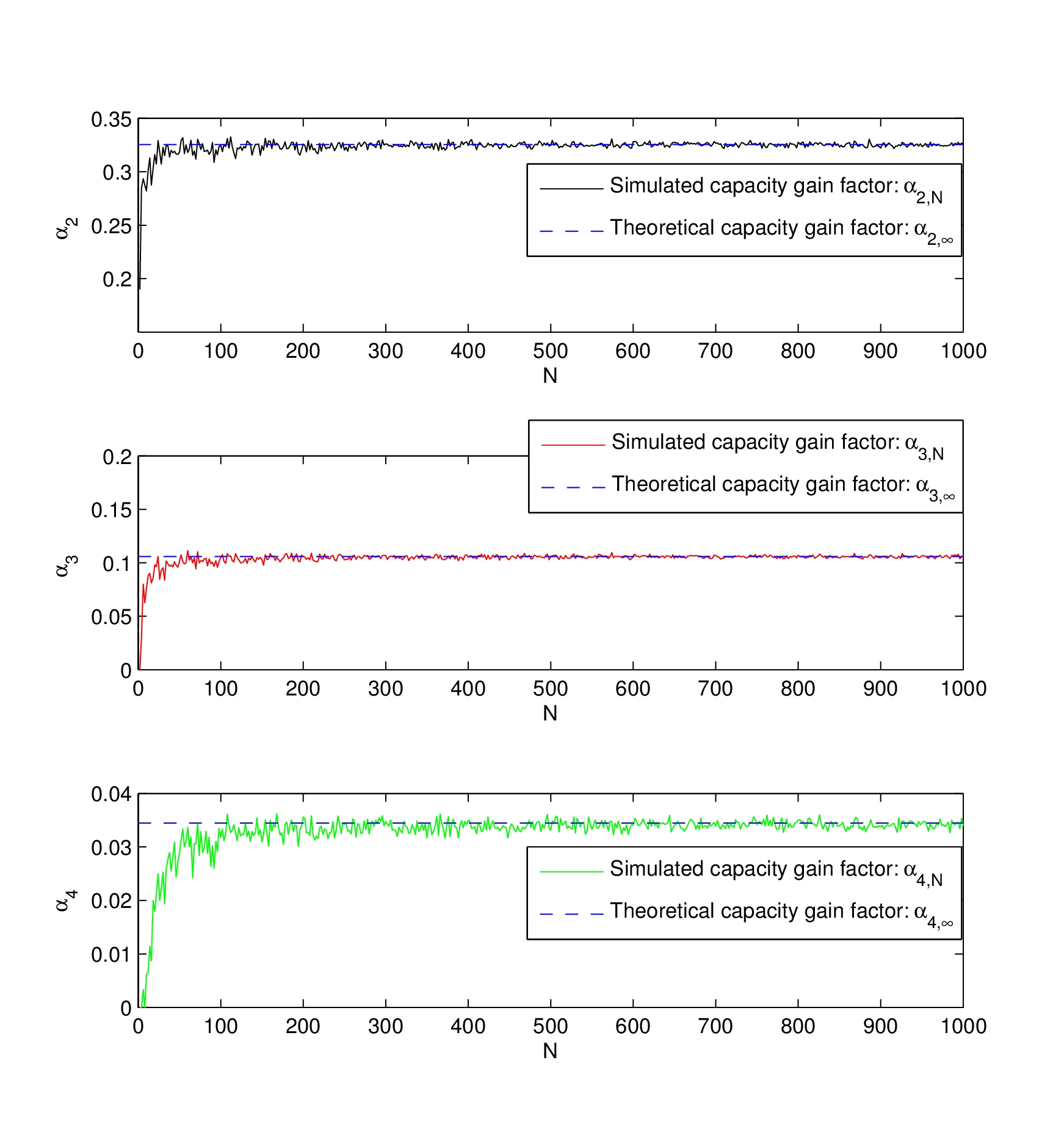}}
\vspace{0cm} \caption{Convergence of band factor gains at SNR = 0 dB.}
\label{fig:cap4}\end{center}
\end{figure}

In \cite{ieeeIT.Majed2}, we analyzed the different configurations of
the sum spectral efficiency for a system with 5 users as function of
the SNR. We showed that at low SNR region, the spectral efficiency
is significantly increased with respect to the traditional system
without cognition while, at high \textrm{SNR} regime, the maximum
sum spectral efficiency reaches $C_{1,\infty}$. In this paper however, we will
focus on the sum spectral efficiency gains as function of $E_b/N_0$.
In fact, the $E_b/N_0$ versus spectral efficiency characteristic is
of primary importance in the study of the behavior of the required
power in the wideband limit (where the spectral efficiency is
small). The key idea behind doing so is to find the best tradeoff
between transmitted energy per information bit and spectral
efficiency \cite{ieeeIT.Verdu}. It is also useful for the sake of
comparing results obtained for different configurations to represent
the fundamental limits in terms of received energy per information
bit rather than the Signal-to-noise ratio.
%As we will see,...
By replacing the SNR in (\ref{eq:hcaror12}) by its equivalent expression in terms of  $E_b/N_0$, the
spectral efficiency of the primary user becomes:
\begin{equation}
C_{1,\infty} \begin{array}[t]{l}
 = \displaystyle
\frac{1}{\ln(2)}\cdot E_i\left({\frac{\gamma_{0}}{\frac{E_b}{N_0}\cdot C_{1,\infty}}
}\right)
\end{array}\,\,
\label{eq:hcaror142}\end{equation} In such a case, the explicit
solution of the spectral efficiency versus $E_b/N_0$ is not
feasible.
In Figure \ref{fig:cap5}, we plot the sum spectral
efficiency gains (with respect to the configuration where only the primary user enjoys the entire band) as function of $E_b/N_0$ where solutions
are given by the implicit equation in (\ref{eq:hcaror142}). The goal here is rather to quantify the spectrum pooling spectral efficiency gain from user to user.
Simulation results were obtained through dichotomic algorithms in Figure \ref{fig:cap5}. We found out that the maximum spectral efficiency gain can not
exceed the range of 60\% for a configuration with one primary user and 4 cognitive users. Notice that, as $E_b/N_0$ increases, all the
configurations tend towards the configuration where only the primary user enjoys the entire band. This can be justified by the fact that,
at high $E_b/N_0$ regime, the water-level $\displaystyle \frac{1}{\gamma_0}$
is becoming greater than the quantity $\displaystyle \frac{N_0}{\left|h\right|^{2}}$ and more power is poured within each  sub-band (see
equation(\ref{eq:snrprob43})).
\\\\
To
proceed further with the analysis, we resort to performance comparison of the proposed scheme with respect to a traditional system where no cognition is used. As far as sum spectral efficiency comparison is concerned, this can be
conducted by considering the two following configurations:
\begin{itemize}
  \item \emph{the non-cognitive radio configuration (NCR)}: where the primary user enjoys the entire bandwidth following an average power constraint per sub-band given by:
  \begin{equation}
\frac{1}{N}\sum_{i=1}^{N} P_{l}^i = L \cdot \overline{P}
\label{eq:hcarr1545}\end{equation}
where $L$ is the maximum number of users at each SNR (as shown in Fig. \ref{fig:cap1}). The primary user can accordingly distribute ($N\cdot L \cdot \overline{P}$) over the $N$ sub-bands in order to maximize his capacity,
  \item \emph{the cognitive configuration}: where ($L-1$) cognitive users coexist with the primary user while sharing the $N$ sub-bands available. Each user has to maximize his capacity with respect to the average power constraint per band of ($card(\Omega_l)\cdot\overline{P}$) as in (\ref{eq:snrprob124}).
\end{itemize}
Figure \ref{fig:cap12} validates the
expectation from the analysis in (\ref{eq:hcaror213}). It clearly shows that the spectrum pooling strategy performs always better than traditional communication system using the same spectral resources due to the multi-user diversity gain. In particular, the spectrum pooling system achieves 1 bit per second per hertz more
than the NCR system.
Let us now focus on the band factor gains expressions. %We
%remark that: \beq \lim_{N\rightarrow \infty} C_{l,N} =
%C_{1,\infty},\qquad \mathrm{for}\,\,\,\  l\in\left[1,L\right] \eeq
%Therefore, by identifying the spectral efficiency where $N$ is
%finite in (\ref{eq:syfeeq}) to the spectral efficiency in
%(\ref{eq:syfe3q}), we compare
%the associated band factor gains. Respectively, we compare
%$\Delta_{l,N}$ to $\Delta_{l,\infty}$.
So far, we have quantified the spectral efficiency gains of different configurations with five users. Let us now investigate how the simulated spectral efficiency gain (with a finite $N$) converges to the theoretical one (when $N$ is assumed to be infinite).
Let us first write the spectral efficiency of each user $l$ as follows:
\begin{equation}
\Phi_{l,\infty} = \alpha_{l,\infty} \cdot
C_{1,\infty},\qquad\mathrm{for}\,\,\,\  l=1,...,L
\label{eq:syfe88q}\end{equation}
where:
\begin{equation}
\alpha_{l,\infty} = \Delta_{\infty}^{l-1} ,\qquad\mathrm{for}\,\,\,\  l=1,...,L
\label{eq:syfe8q}\end{equation}
Note here that $\alpha_{l,\infty}$ represents the band factor gain from the primary user to user $l$. In Figure \ref{fig:cap4}, numerical simulation is carried out by considering a system with four cognitive users. We compared simulated values of $\alpha_{l,N}$ based on equation (\ref{eq:syfe7q}) to theoretical values in (\ref{eq:syfe8q})
for each user $l$ and for SNR = 10 dB. We remark that as $N$
increases, the simulated band factor gain tends to $\alpha_{l,\infty}$. Moreover,
simulation results show that $\alpha_{2,N}$ converges more rapidly
to the associated theoretical gain factor value than for user 3 or
user 4.
\vspace{2cm}
\section{Conclusion}

In this work, we have considered a new strategy called Spectrum Pooling enabling public access to the new spectral ranges without sacrificing the transmission quality of the actual license owners. For the first time, our analysis has quantified the achievable gain of
using spectrum pooling with respect to classical radio devices. We found out that though its simplicity, the proposed scheme is effective to provide a higher spectral efficiency gain than the classical scheme does. We further obtained a characterization of the achievable spectral
efficiency as well as the maximum number of possible pairwise
communications within such a scenario. Simulation results validate our theoretical claims and offer insights into how much one can gain from spectrum pooling in terms of spectral efficiency.
As a future work, it is of major interest to generalize the
problem to limited feedback in order to characterize the
sum spectral efficiency gain of such cognitive protocols with respect to
the proposed scenario. It would be further interesting to measure the throughput of the proposed protocol given a realistic primary system model (e.g., ethernet traffic) compared to an OFDM/TDD overlay cognitive radio system.

\pagebreak

\pagebreak


\pagebreak

\end{document}